# Application for Evaluation of the Professional Competencies of the Teaching Staff

**Stud. Ovidiu Crista,**
**Assoc.Prof. Tiberiu Marius Karnyanszky, Ph.D.**
**“Tibiscus” University of Timişoara, România**

ABSTRACT. The goal of the presented application is to offer a full support for universities in retrieving the feedback from their students with regard to their teachers. This is the main reason we described it in this paper.
To build this application the following tools have been used: Microsoft Notepad 5.1 (to make the source files), Adobe Photoshop CS3, (to make the background image), and Adobe Flash Media Encoder 8 (to render the video clips).

## 1. Generalities

**To run the application the following tools and technologies should be used:**

1. a computer (any platform) which runs a web server with PHP: Hypertext Preprocessor installation, and a MySQL server;
2. a computer (any platform) with a web browser;

The computers mentioned above can be one single computer which plays both of the server and client roles. If it is not the case, then an Internet connection would be needed for both of them to ensure a communication.

## 2. Some explanations about the technical side

This application acts like a website which interacts with its users, having a database behind for storing data. The source files are stored to a virtual directory on the web server and a specific Uniform Resource Locator is assigned. Each time the application is started, the source files are interpreted by the PHP and finally, it sends the HTML output to the web browser as a response to the user's action.

## 3. The interfaces

### 3. The student interface

This is the web page which communicates with the students by displaying the questions about their teachers, and waiting for their responses; this communication is possible through the use of the HTML forms (fig. 1). The URL of the student interface never changes, so a single physical file can be considered as performing all the tasks. The output is still different each time a step forward has been made. The submitted values and the recordings from the database help the application to know exactly what content to deliver at one time. Multiple php files on that virtual directory are involved to make the output. Because the application must behave differently for each student, the Internet Protocol address is registered so all the values of the variables depend on that address which identify a certain student filling in the questionnaire.

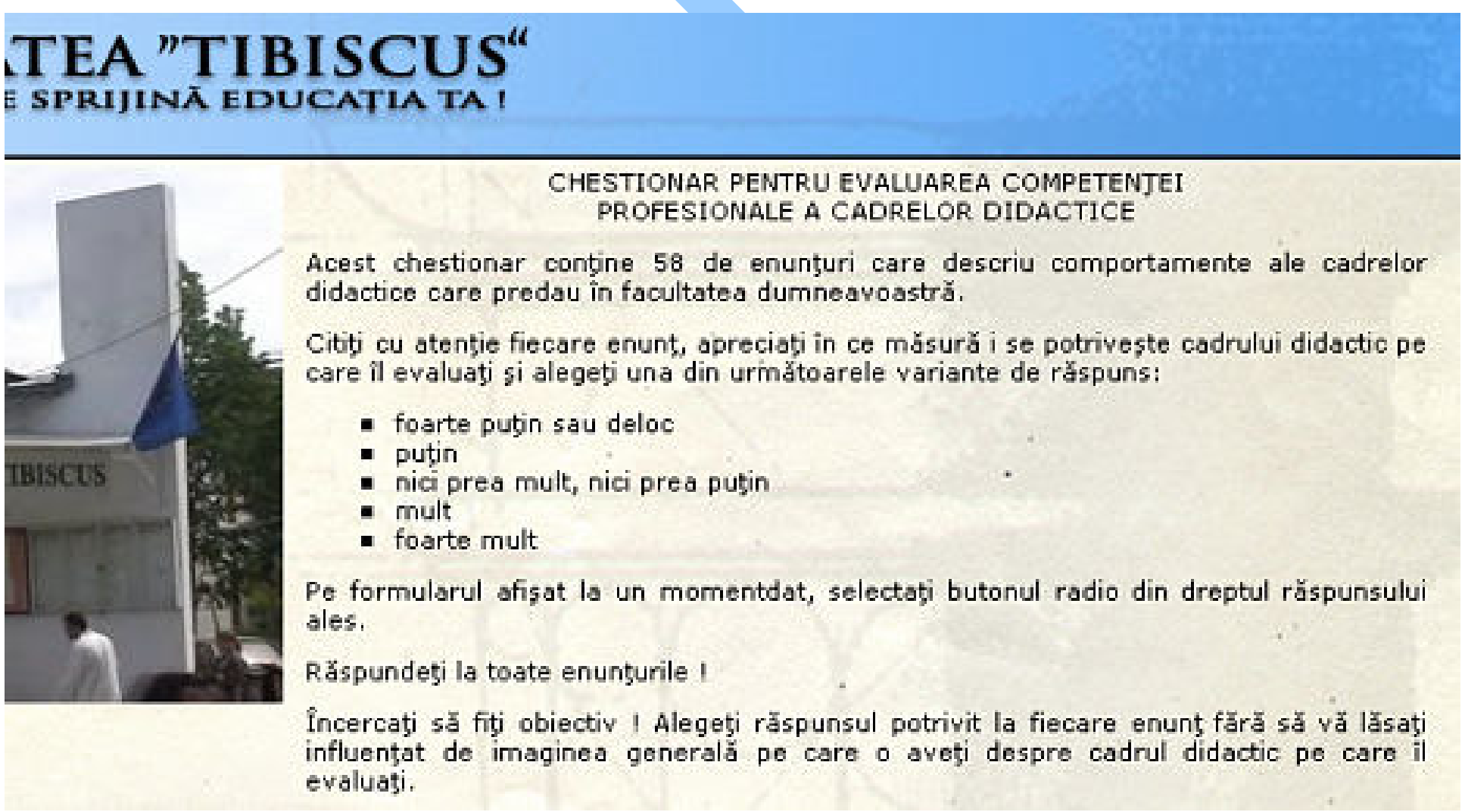

*Figure 1.*

Each time a student answers a question, the corresponding value of his or her answer together with the number of the last answered question are stored in the database. This makes sure that the student could not go back to the already answered questions or skip some of them; the application would only accept and save answers to the questions that have to be answered.

The student interface can be in three different states:

- when the application is activated (as explained later) and it is accessed from a predefined list of agreed IP addresses;
- when the application is activated and it is accessed from any other IP address;
- when the application is not activated and it is accessed from any location;

One student would be permitted to answer the questions specific to one teacher only once, so the final results cannot be misled. The first case represents the expected completion when the set of the answers is considered for the evaluation.

The two other cases represent the testing version of the application, when a Reset button is provided so the user can delete all recordings corresponding to his or her IP address, and then start from the beginning without compromising the needed information.

For all the cases, there is a same set of source file which compare the user's IP address to the ones from the agreed list.

The web page representing the student interface is composed by the background image, the flash video clip and the HTML form. Even if the question that has to be answered is displayed as text, to avoid monotony, that question is asked the same time using an embedded flash video clip.

The HTML form contains the text of the question, the five radio buttons for each possible answer and the submit button (fig. 2). Occasionally, a status message may appear, if something was wrong. None of the radio buttons are selected by default. The user has to click one of them to indicate the answer and then click the submit button. Otherwise, the same question and an error message will be displayed.

Though probably nobody will try to fool somehow the application, some situations where considered to avoid answering again to an already answered question or trying to skip some to get to the end faster. The browser's back button can help to retrieve old version (from the cache) of the web page. No matters how the initial form is made (from the browser's cache or from an edited version stored locally on the user's computer) when the form results are sent back to the web server, the application, knowing which is the last question that user really answered, performs a test on what question has been answered this time.

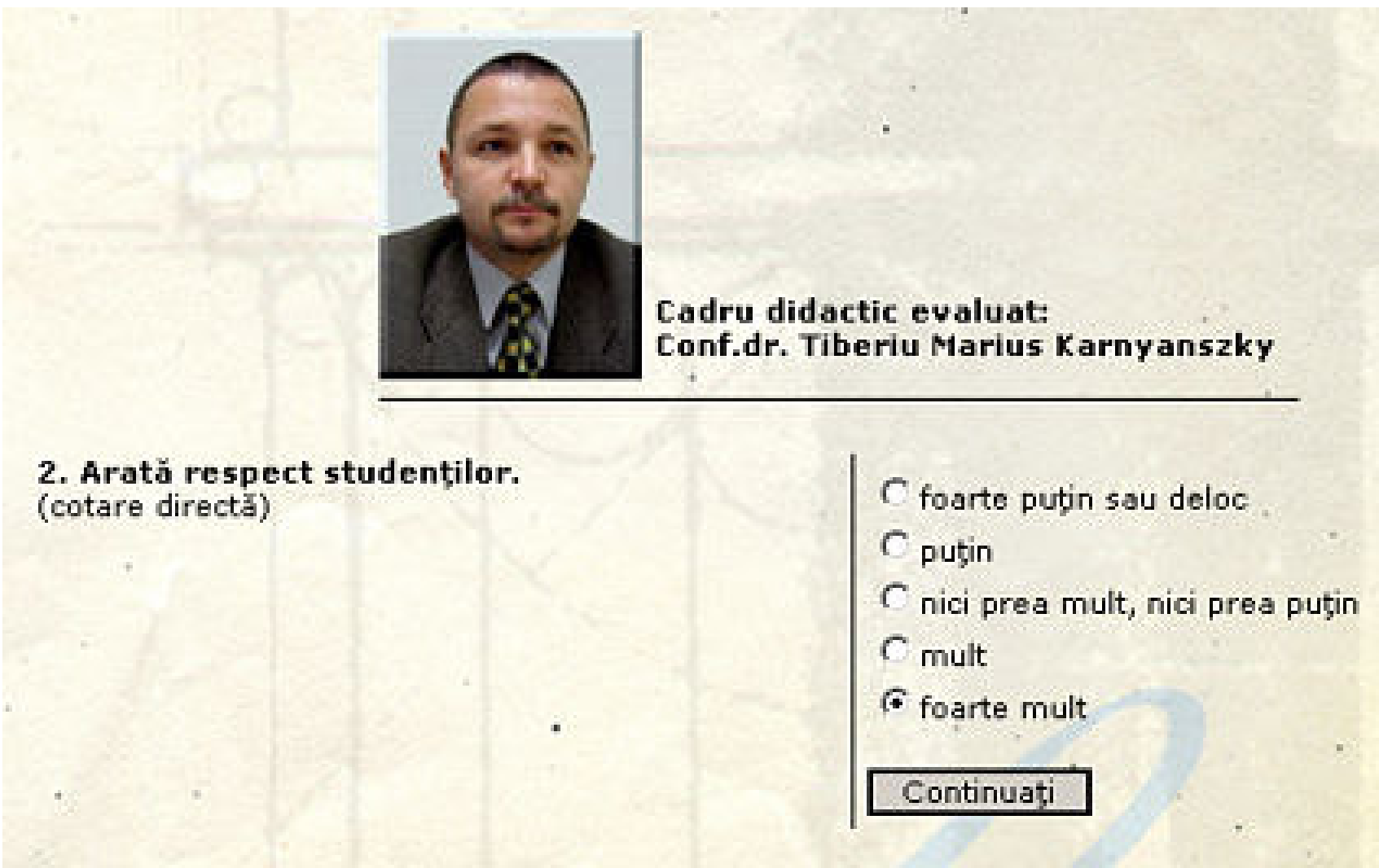

*Figure 2.*

### 3.2. The admin interface

An administration panel is also provided in order to assist the administrator to set one of those states, and also to set other variables. Access to this interface is permitted from any location world wide only by providing the correct username and password (fig. 3).

There are three sections:

- view status

This offers the administrator a general information about the database, the status, the number of people filling in the questionnaire, the locations of them (without information about their answers) and the teacher currently evaluated;

- modification of parameters

Here is where the administrator can activate/deactivate the application, set the teacher to be evaluated and create or modify the list of the agreed IP's;

- teaching staff

This section contains the list of the teachers that can be selected for evaluation. The administrator can modify this list or add new information on other teachers;

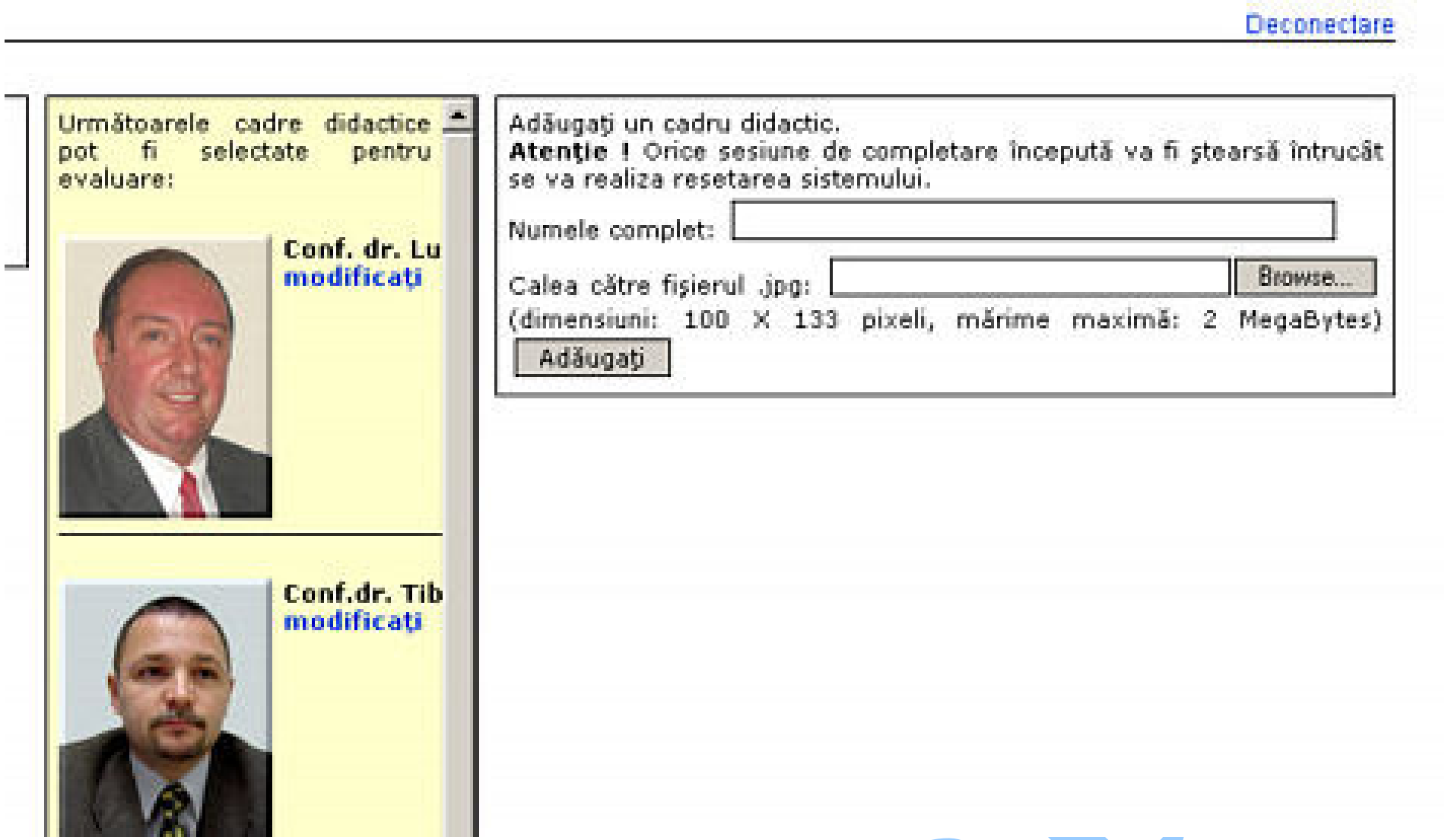

*Figure 3.*

## 3.3.The results interface

All the answers are stored in the database. This interface (fig. 4) lists the set of the answers for every teacher that was evaluated. It can also prepare a complete page for printing results.

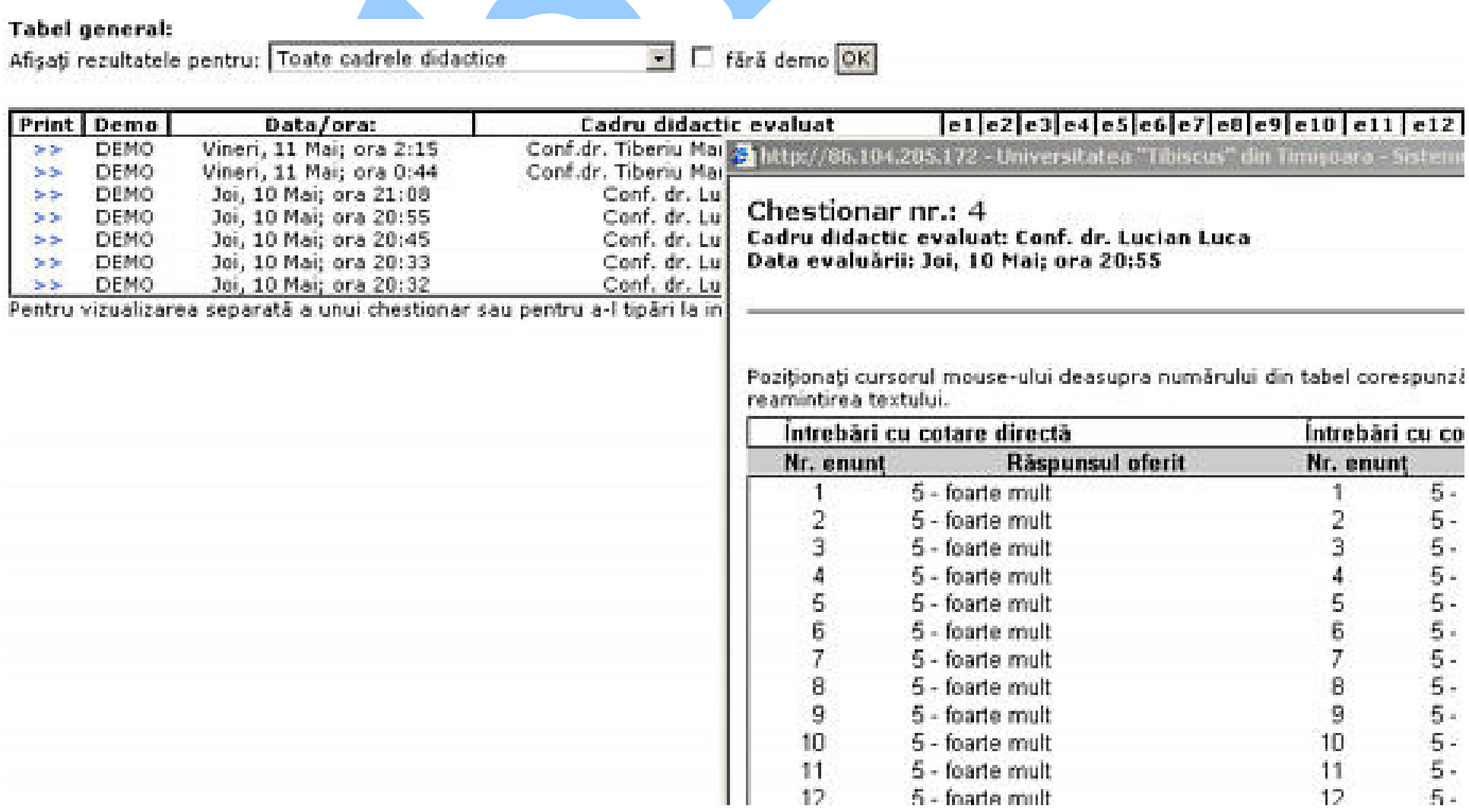

*Figure 4.*

## 4. Conclusions and improvements for the future

All the features offered now are enough for storing and retrieving data. For the future, we plan to program more features for the admin interface and the results. Next step might be the elaboration of statistics on the results. We look forward to implement a maximum time for completion or a maximum time for answering one question; a response through the video clip asking the user to answer if he or she is getting late.